\documentclass[final,12pt, times]{elsarticle}
\makeatletter
\def\ps@pprintTitle{%
 \let\@oddhead\@empty
 \let\@evenhead\@empty
 \let\@oddfoot\@empty
 \let\@evenfoot\@empty
}
\makeatother

\usepackage{amssymb}
\usepackage{graphicx}
\usepackage{dcolumn}
\usepackage{amsthm}
\usepackage{float}
\usepackage{xcolor}
\usepackage[caption = false]{subfig}
\usepackage{graphicx}
\usepackage{varwidth}
\usepackage{amsmath}
\usepackage{bm}
\usepackage{siunitx}
\usepackage[T1]{fontenc}
\usepackage{hyperref}
\usepackage{cleveref}

\begin{document}

\begin{frontmatter}



\title{Splay Stiffening and Twist Softening in a Ferroelectric Nematic Liquid Crystal}

\author[label1,label2]{Evangelia E. Zavvou\corref{cor1}}
\author[label2]{Alexander Jarosik}
\author[label2]{Hajnalka N\'adasi}
\author[label1]{Christoforos A. Krontiras}
\author[label1]{Panagiota K. Karahaliou}
\author[label3]{Rachel Tuffin}
\author[label3]{Melanie Klasen-Memmer}
\author[label2]{Alexey Eremin \corref{cor1}}

\affiliation[label1]{organization={Department of Physics, University of Patras},
            city={Patras},
            postcode={26504},
            country={Greece}}
            
\affiliation[label2]{organization={Otto von Guericke University Magdeburg, Institute of Physics, Dept. Nonlinear Phenomena},
            city={Magdeburg},
            postcode={39106},
            country={Germany}}

\affiliation[label3]{organization={Electronics Division, Merck KGaA},
            city={Darmstadt},
            postcode={64293},
            country={Germany}}
\cortext[cor1]{Corresponding authors. E-mail addresses:\newline
\href{alexey.eremin@ovgu.de}{alexey.eremin(at)ovgu.de} (A. Eremin),\newline \href{ezavvou@upatras.gr}{ezavvou(at)upatras.gr} (E. Zavvou).}

\begin{abstract}
The recent discovery of ferroelectric nematics—genuine 3D ferroelectric fluids—has underscored the importance of electrostatic interactions in shaping the physical behaviour of soft matter systems. In this paper, we investigate the mechanical properties of ferroelectric nematics by directly comparing the splay and twist elastic constants in a liquid crystal system that exhibits both nonpolar and ferroelectric nematic phases. Our results reveal that polar ordering results in increased splay rigidity and a concomitant reduction in twist elasticity.
\end{abstract}

%
%
\begin{keyword}
Liquid Crystals \sep Ferroelectric Nematic Phase \sep Fr\'eedericksz Transition \sep Elastic Constants \sep Dielectric Relaxation \sep Nonlinear Optics


\end{keyword}

\end{frontmatter}



\section{\label{sec:intro} Introduction}
Uniaxial nematic liquid crystals constitute the simplest and most widely studied class of liquid crystalline phases due to their fundamental importance and diverse applications~\cite{deGennes:1995vg}. The nematic phase (N) is characterised by long-range orientational order, with molecules aligning on average along a common axis, defined by the nematic director $\mathbf{n}$. However, the symmetry of conventional nematics is not polar but cylindrical, as reflected in the invariance of the director under inversion ($\mathbf{n} \to -\mathbf{n}$). This symmetry fundamentally dictates the mechanical properties of nematics, particularly their response to director deformations. 

A unique property of liquid crystals is their ability to transmit torques through deformations of the director field ~\cite{deGennes:1995vg}. In 1927, Fr\'eedericksz demonstrated that when a nematic liquid crystal confined between two parallel plates is exposed to external electric or magnetic field, the director reorientation—known as the Fr\'eedericksz transition—exhibits a threshold behaviour~\cite{Freedericksz.1927}. By analysing changes in optical transmittance or capacitance during this transition, one can determine the nematic elastic constants~\cite{deGennes:1995vg}.

Nematic phases with vector symmetries have recently been identified in both ferromagnetic and ferroelectric systems. In 2017, spontaneous polar order in a nematic phase composed of low-molecular-mass mesogens was reported almost simultaneously by Mandle et al. and Nishikawa et al. ~\cite{Nishikawa.2017,mandle2017rational,Mertelj.2018, Chen.2020}. The key mesogens exhibiting the ferroelectric nematic phase (N$_{\mathrm{F}}$) are RM734 and DIO, which are characterised by a wedge-like molecular shape and multiple dipolar groups. These features result in an overall longitudinal dipole moment on the order of 10 D, along with strong optical nonlinearity and an exceptionally high dielectric permittivity in thin-film cells ~\cite{Brown.2021ysv,Sebastian.2020}. The precise nature of this dielectric response remains an active area of research, with the Polarisation Capacitance Goldstone (PCG) mode and Continuous Phenomenological Model (CPM) being among the leading theoretical frameworks under experimental investigation~\cite{Clark.2024,Adaka.2024,Vaupotic.2023,Erkoreka.2023}.

Research on ferroelectric nematics is driven by their potential applications in fast electrooptical and electromechanical devices. The mechanoelectrical effect in the N$_{\mathrm{F}}$ phase~\cite{Rupnik.2024} opens new possibilities for sensor and touch applications, as well as tunable optical filters. Additionally, their strong optical nonlinearity makes N$_{\mathrm{F}}$ materials promising candidates for photonic devices and communication technologies. 

Over the past few years, intensive research efforts have led to a rapid expansion of the catalogue of mesogens capable of forming the N$_{\mathrm{F}}$ phase. Alongside these developments, there has been substantial progress in characterizing the $\mathrm{N}$–N$_{\mathrm{F}}$ and N–M–N$_{\mathrm{F}}$ phase transitions, where M denotes an intermediate antiferroelectric phase~\cite{zattarin2024design, Mrukiewicz.2024}. This phase has been proposed to correspond to the antiferroelectric SmZ$_A$ phase, in which the nematic director is oriented perpendicular to the layer normal, leading to distinct structural and dielectric properties~\cite{Chen.2023}. Antiferroelectric order in this phase can be manipulated by the addition of ionic liquids as demonstrated by Rupnik et al. ~\cite{medle2025antiferroelectric}.

Polar heliconical phases have also been discovered, in which the helical pitch can be controlled by small electric fields~\cite{Nishikawa.2021,Gorecka2024,Gibb.2024}. This allows for tuning of the selective reflection in the visible range.

The coupling between the spontaneous electric polarisation and the nematic director results in an electrostatic contribution to the director's deformation energy, significantly affecting the material's mechanical properties. The avoidance of the polarisation charges resulting from the divergence of the director stiffens the splay elasticity of the N$_{\mathrm{F}}$ phase~\cite{Kumari.2023,Basnet.2022bek,Caimi.2023t5j,Chen.2020,Zavvou.2022gmk}. Such stiffening is reflected in the formation of conics in thin N$_{\mathrm{F}}$ films on a glycerine surface and also in the drastic increase of the threshold of the magnetic Fr\'eederiksz transition reported for the N$_{\mathrm{F}}$ compounds~\cite{Kumari.2023,Zavvou.2022gmk}. In our previous paper, we reported on the measurements of the splay and the twist elastic constants in a compound exhibiting a direct Iso - N$_{\mathrm{F}}$ phase transition~\cite{Zavvou.2022gmk}. 

In this paper, we explore the mechanical, dielectric and nonlinear optical properties of a liquid crystalline mixture exhibiting the intermediate antiferroelectric phase in between the nematic and the ferroelectric nematic phases, with the latter being stable at and below room-temperature. We directly compare the elastic constants of the nonpolar N and the polar N$_{\mathrm{F}}$ phases through the analysis of the capacitance and/or optical transmission recorded during the Fr\'eedericksz transition. The temperature dependence of the diamagnetic susceptibility in the N phase is determined through complementary studies in external electric and magnetic fields. To avoid the undesired electric field effects, the mechanical properties within the N$_{\mathrm{F}}$ phase are explored through measurements of the optical transmission in an external magnetic field.

\section{\label{sec:experimental} Experimental}

The studied material, MDA-21-671 (provided by Merck Electronics KGaA, Darmstadt, Germany), is a mixture exhibiting the following phase sequence: Crystal $< -20^\circ$C $\rightarrow$ N$_\mathrm{F}$ 45.8 $^\circ$C $\rightarrow$ M 57.9 $^\circ$C $\rightarrow$ N 87.6 $^\circ$C $\rightarrow$ Isotropic. The coexistence range of the nematic and isotropic phases is narrow, not exceeding $2^\circ$C~\cite{jarosik2024fluid}.

Phase identification and birefringence measurements were carried out using an AxioImager A.1 polarising optical microscope (Carl Zeiss, GmbH, Germany) equipped with a heating stage (Instec, USA) and a Berek tilting compensator. The material was filled into a planar-aligned cell (E.H.C., Japan) with a $\qty{6}{\micro\meter}$ gap and parallel rubbing. Measurements were taken upon cooling from the isotropic phase using monochromatic light with wavelength $\lambda = \qty{546}{nm}$.

The presence of polar order in the liquid crystalline phases of the studied compound was confirmed via optical second harmonic generation (SHG) measurements using a Nd:YAG laser. The fundamental beam had a wavelength of $\lambda = \qty{1064}{nm}$, a pulse width of 10 ns, and a repetition rate of \qty{10}{\hertz}. The sample was introduced into custom-made in-plane-switching (IPS) cells with planar alignment and a thickness of \qty{10}{\micro\meter}. The incident beam struck the sample at an angle of $30^\circ$ to the cell normal. The SHG signal was collected in transmission using a photomultiplier tube (Hamamatsu) and calibrated against a \qty{50}{\milli\meter} reference quartz plate. Measurements were carried out upon cooling from the isotropic phase in $1^\circ$C increments.

Fluorescence and SHG microscopy studies were performed using a multiphoton (MP) confocal microscope (Leica TCS SP8) with excitation at the fundamental wavelength $\lambda = \qty{880}{\nano\meter}$. For fluorescence imaging, samples were doped with 0.01 wt$\%$ of the dichroic dye N,N’-bis(2,5-di-tert-butylphenyl)-3,4,9,10-perylenedicarboximide (BTBP, Sigma-Aldrich) to visualize the director orientation. Excitation was performed using a laser source at $\lambda_{\mathrm{ex}} = \qty{488}{nm}$.

The complex dielectric permittivity, $\varepsilon^*(\omega) = \varepsilon'(\omega) - i \varepsilon''(\omega)$, in the isotropic and liquid crystalline phases of the studied mixture was measured over the frequency range from 0.1 Hz to 1 GHz. This was achieved by combining a Novocontrol Alpha-N Frequency Response Analyser ($f = \qty{0.1}{\hertz} - \qty{1}{\mega\hertz}$) for the low-frequency regime and an HP4291B RF Impedance Analyser ($f = \qty{0.1}{\mega\hertz} - \qty{1}{\giga\hertz}$) for the high-frequency range.

The sample was sandwiched between two untreated, gold-plated brass electrodes (\qty{5}{\milli\meter} diameter), separated by \qty{50}{\micro\meter}-thick silica fibres. For low-frequency measurements, the cell was mounted in a modified BDS1200 sample holder; in the high-frequency regime, a BDS2200 RF sample cell was used, placed at the end of a coaxial line (Novocontrol). In both cases, the sample cell was housed in a Novocontrol cryostat, with temperature precisely controlled and stabilized to within $\pm \qty{0.02}{\celsius}$ using a Quatro Cryosystem (Novocontrol).

Isothermal frequency scans were recorded upon slow cooling from the isotropic phase at a rate of $\qty{0.25}{\celsius\per\minute}$, with variable temperature steps. The oscillating voltage was set to $U_{\mathrm{rms}} = 0.01 \mathrm{V}$. Data acquisition and storage were managed via WinDETA software (Novocontrol).

The resulting dielectric spectra were analysed using the Havriliak–Negami (HN) equation:

\begin{equation}
\varepsilon^*(\omega) = \varepsilon_\infty + \sum_{\kappa} \frac{\Delta \varepsilon_{\kappa}}{\left(1 + (i \omega \tau_{\mathrm{HN},\kappa})^{\alpha_\kappa}\right)^{\beta_\kappa}} + \left(\frac{\sigma_{\mathrm{DC}}}{i \omega \varepsilon_0}\right)^{N},
\label{eq:HN}
\end{equation}

\noindent where $\varepsilon_\infty$ is the high-frequency permittivity, $\Delta \varepsilon_{\kappa}$ denotes the dielectric strength of the corresponding relaxation process, $\sigma_{\mathrm{DC}}$ is the DC conductivity, and $N$ is an exponent with values between 0 and 1. The characteristic relaxation time $\tau_{\mathrm{HN},\kappa}$ is related to the frequency of maximum dielectric loss via:
\begin{equation}
f_{\mathrm{max},\kappa} = 
\frac{1}{2\pi \tau_{\mathrm{HN},\kappa}} \left[ \sin\left( \frac{\pi \alpha_\kappa}{2+2\beta_\kappa} \right) \right]^{\frac{1}{\alpha_\kappa}} 
\left[ \sin\left( \frac{\pi \alpha_\kappa \beta_\kappa}{2+2\beta_\kappa} \right) \right]^{-\frac{1}{\alpha_\kappa}},
\label{eq:tau-max}
\end{equation}

The shape parameters $\alpha_\kappa$ and $\beta_\kappa$ characterise the symmetric and asymmetric broadening of the relaxation time distribution, respectively, and satisfy the condition $0 < \alpha_\kappa$, $\alpha_\kappa \beta_\kappa \leq 1$. When $\alpha_\kappa = \beta_\kappa = 1$, Eq.~\eqref{eq:HN} reduces to the Debye equation. Deconvolution of the dielectric spectra was performed using the WinFit software (Novocontrol).
 
The mechanical and magnetooptical properties were evaluated through complementary magnetic and electric Fr\'eedericksz transition (FT) experiments. In both cases, the material was filled into planar cells (E.H.C., Japan) with a thickness of \qty{25}{\micro\meter}, equipped with parallel rubbing alignment layers and ITO electrodes with a sheet resistance of $\qty{10}{\ohm}$.

For the electric Fr\'eedericksz transition, the cell capacitance was measured as a function of the applied alternating voltage ($U_{\mathrm{rms}} = 0.01$–$\qty{3}{\volt}$) using an Alpha-N Analyser (Novocontrol). A time interval of \SI{30}{\second} was maintained between the application of the electric field and the capacitance measurement. The frequency of the applied AC field ($f = 0.7$–$\qty{5}{\kilo\hertz}$) was carefully optimised at each temperature to minimise low-frequency ionic effects, while ensuring the measurement remained within the quasi-static dielectric regime. To achieve an undistorted initial director configuration, the sample was heated to the isotropic phase between successive FT measurements and then slowly cooled to the target temperature.

The measured dielectric permittivity as a function of applied AC voltage, $\varepsilon_{\mathrm{EQ}}(U)$, was interpreted as the effective permittivity of a series capacitor network consisting of three layers: two polyimide (PI) alignment layers with capacitance $C_{\mathrm{PI}} = \varepsilon_{\mathrm{PI}} \varepsilon_0 A / d_{\mathrm{PI}}$, and the liquid crystal (LC) layer with capacitance $C_{\mathrm{LC}} = \varepsilon_{\mathrm{LC}} \varepsilon_0 A / d$. Here, $A$ is the electrode area, $d$ and $d_{\mathrm{PI}}$ are the thicknesses of the LC and PI layers, respectively. The empty cell capacitance was defined as $C_0 = \varepsilon_0 A / d$ and was measured prior to sample filling.
The dielectric permittivity of the LC layer, $\varepsilon_{\mathrm{LC}}(U)$, was recalculated from the measured $\varepsilon_{\mathrm{EQ}}(U)$ using the relation: $\varepsilon_{\mathrm{LC}} = \varepsilon_{\mathrm{EQ}} \varepsilon_{\mathrm{PI}}/\left[\varepsilon_{\mathrm{PI}} - 2\varepsilon_{\mathrm{EQ}}(d_{\mathrm{PI}}/d)\right]$. In this calculation, the PI layer thickness was assumed to be $d_{\mathrm{PI}} = \qty{20}{\nano\meter}$ (LX-1400 polyimide by Hitachi), and the permittivity of the polyimide was taken as $\varepsilon_{\mathrm{PI}} = 3.5$, a typical value that is expected to be nearly temperature- and frequency-independent.

The dielectric anisotropy and the splay ($K_{11}$) and bend ($K_{33}$) elastic constants were determined in the high-temperature nematic phase by fitting the recalculated $\varepsilon_{\mathrm{LC}}(U)$ curves using the expressions:
\begin{equation}
       U=\frac{2 U_{\mathrm{th}}}{\pi} \sqrt{1+\gamma \sin^2\theta_{\mathrm{m}}} 
       \int_{0}^{\pi / 2}\left[\frac{1+\kappa \sin^2\theta_{\mathrm{m}} \sin ^2 \psi}{\left(1+\gamma \sin^2\theta_{\mathrm{m}} \sin ^2 \psi\right)\left(1-\sin^2\theta_{\mathrm{m}} \sin ^2 \psi\right)}\right]^{1 / 2} d \psi,
 \label{eq:DEU_1} 
\end{equation}

and

\begin{equation}
    \varepsilon=\varepsilon_{\perp} \frac{\int_{0}^{\pi / 2}\left[\frac{\left(1+\gamma \sin^2\theta_{\mathrm{m}} \sin ^2 \psi\right)\left(1+\kappa \sin^2\theta_{\mathrm{m}} \sin ^2 \psi\right)}{\left(1-\sin^2\theta_{\mathrm{m}} \sin ^2 \psi\right)}\right]^{1 / 2} d \psi}{\int_{0}^{\pi / 2}\left[\frac{1+\kappa \sin^2\theta_{\mathrm{m}} \sin ^2 \psi}{\left(1+\gamma \sin^2\theta_{\mathrm{m}} \sin ^2 \psi\right)\left(1-\sin^2\theta_{\mathrm{m}} \sin ^2 \psi\right)}\right]^{1 / 2} d \psi},
 \label{eq:DEU_2} 
\end{equation}
where, $\theta_{\mathrm{m}}$ is the director deflection angle in the mid-plane, $\kappa=(K_{33}-K_{11})/K_{11}$ and $\gamma=(\varepsilon_{\parallel}-\varepsilon_{\perp})/\varepsilon_{\perp}$.

For the magnetic Fr\'eedericksz transition, \qty{25}{\micro\meter}-thick planar cells filled with the material were placed between a pair of Helmholtz coils capable of generating magnetic flux densities up to \qty{650}{\milli\tesla} . The magnetic field–induced splay and twist Fréedericksz transitions enabled the determination of the corresponding elastic constants, $K_{11}$ and $K_{22}$, by monitoring the optical transmission of the sample between crossed polarising prisms (Thorlabs) as a function of the applied magnetic field.

The sample was illuminated using a He–Ne laser. The rubbing direction of the cell was oriented perpendicular to the magnetic field and at an angle of $45^\circ$ with respect to the crossed polarisers to maximize birefringence sensitivity.

In the splay geometry, the magnetic field was applied perpendicular to the cell plane; in the twist configuration, it was applied parallel to the cell plane. All measurements were carried out upon cooling from the isotropic phase.

\section{\label{sec:results} Results and Discussion}

\subsection{\label{sec:POM} Optical Characterisation}

\begin{figure}[ht!]
\centering
 \includegraphics[width=0.7\columnwidth]{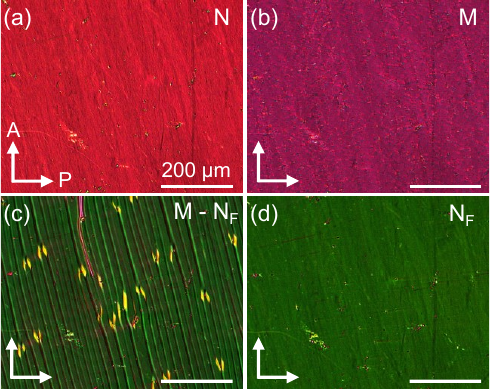}
  \caption{Polarising optical microscopy textures of the studied compound in a \qty{6}{\micro\meter} thick planar cell with parallel rubbing: (a) N phase at \qty{60}{\celsius} , (b) M phase at \qty{46.4}{\celsius}, (c) N$_{\mathrm{F}}$ phase at \qty{38.8}{\celsius} and (d) N$_{\mathrm{F}}$ phase at \qty{28}{\celsius}. The rubbing direction forms an angle of \qty{15}{\degree} with the analyser. The white bar is \qty{200}{\micro\meter}.}
  \label{fig:POM}
\end{figure}

Polarising optical microscopy (POM) observations and birefringence measurements were performed using commercial planar cells with parallel rubbing and a spacing of \qty{6}{\micro\meter}. Upon cooling from the isotropic liquid, a uniformly aligned nematic  phase emerged, as shown in Fig.~\ref{fig:POM}a.

The transition to the intermediate M phase was accompanied by a progressive suppression of director fluctuations (flickering), along with the development of a grainy texture (Fig.~\ref{fig:POM}b). Further cooling led to the formation of the N$_{\mathrm{F}}$ phase, characterised by the appearance of a striped texture, with stripes oriented parallel to the rubbing direction. These observations are consistent with previously reported studies of the mesogen DIO, which exhibits the same phase sequence as the studied mixture~\cite{Nishikawa.2017,sebastian2022ferroelectric}.

At lower temperatures, line defects were observed (Fig.~\ref{fig:POM}c), which subsequently annihilated, giving rise to a uniformly aligned N$_{\mathrm{F}}$ phase. This alignment is promoted by the parallel rubbing of the top and bottom glass substrates~\cite{sebastian2021electrooptics}.

\begin{figure}[ht!]
\centering
\includegraphics[width=0.7\columnwidth]{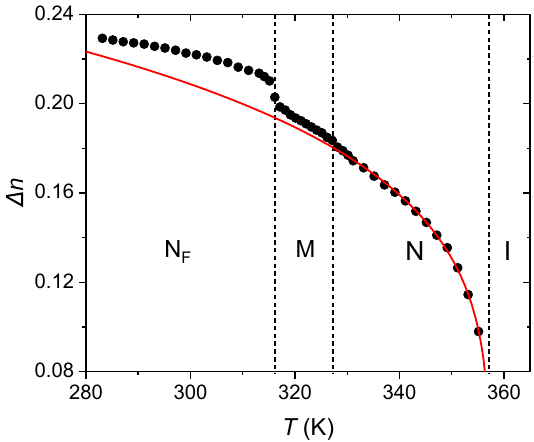}
  \caption{Temperature dependence of birefringence ($\Delta n$) in the liquid crystalline phases of the studied material determined at  $\lambda = \qty{545}{\nano\meter}$. Black circles correspond to experimental data, while the solid line represents the theoretical fitting according to Eq.~\eqref{eq:Haller}.}
  \label{fig:Birefringence}
\end{figure}

The temperature dependence of birefringence, measured upon cooling from the isotropic liquid, is shown in Fig.~\ref{fig:Birefringence}. Within the nematic  phase, the birefringence increases on cooling and follows a Haller-like temperature dependence~\cite{Haller1975}:

\begin{equation}
\Delta n = \Delta n_0 \left(1 - \frac{T}{T_{\mathrm{IN}}^{\ast}}\right)^\beta
\label{eq:Haller}
\end{equation}

\noindent where $\Delta n_0$ is the extrapolated birefringence at $T=\qty{0}{\kelvin}$, $T_{\mathrm{IN}}^{\ast}$ is a fitting parameter slightly above the isotropic–nematic transition temperature, and $\beta$ is a critical exponent. The best-fit parameters are $\Delta n_0 = 0.315 \pm 0.005$, $T_{\mathrm{IN}}^{\ast} = (357.2 \pm 0.3)\mathrm{K}$, and $\beta = 0.225 \pm 0.006$.

The orientational order parameter, defined as $S = \Delta n / \Delta n_0$, reaches values around $S \approx 0.57$ near the N–M phase transition. Notably, both the Haller-like behavior of $\Delta n$ in the high-temperature N phase and the small value of $\beta \sim 0.2$ are characteristic of conventional nematogens, as observed in other ferroelectric liquid crystals~\cite{Chen.2020, Brown.2021ysv, yadav2023spontaneous}.

Upon entering the intermediate M phase, birefringence progressively deviates from the Haller trend, increasing above the expected values. A distinct jump in $\Delta n$ is observed at the M–N$_{\mathrm{F}}$ transition, followed by a gradual increase within the N$_{\mathrm{F}}$ phase. These findings are consistent with the weakly first-order character of the M–N$_{\mathrm{F}}$ transition~\cite{Nishikawa.2017, Brown.2021ysv}, and they further support an increase in orientational order associated with the emergence of long-range polar order in the N$_{\mathrm{F}}$ phase.

\subsection{\label{sec:NLO} Nonlinear Optical Response}
\begin{figure}[ht!]
\centering
\includegraphics[width=0.8\columnwidth]{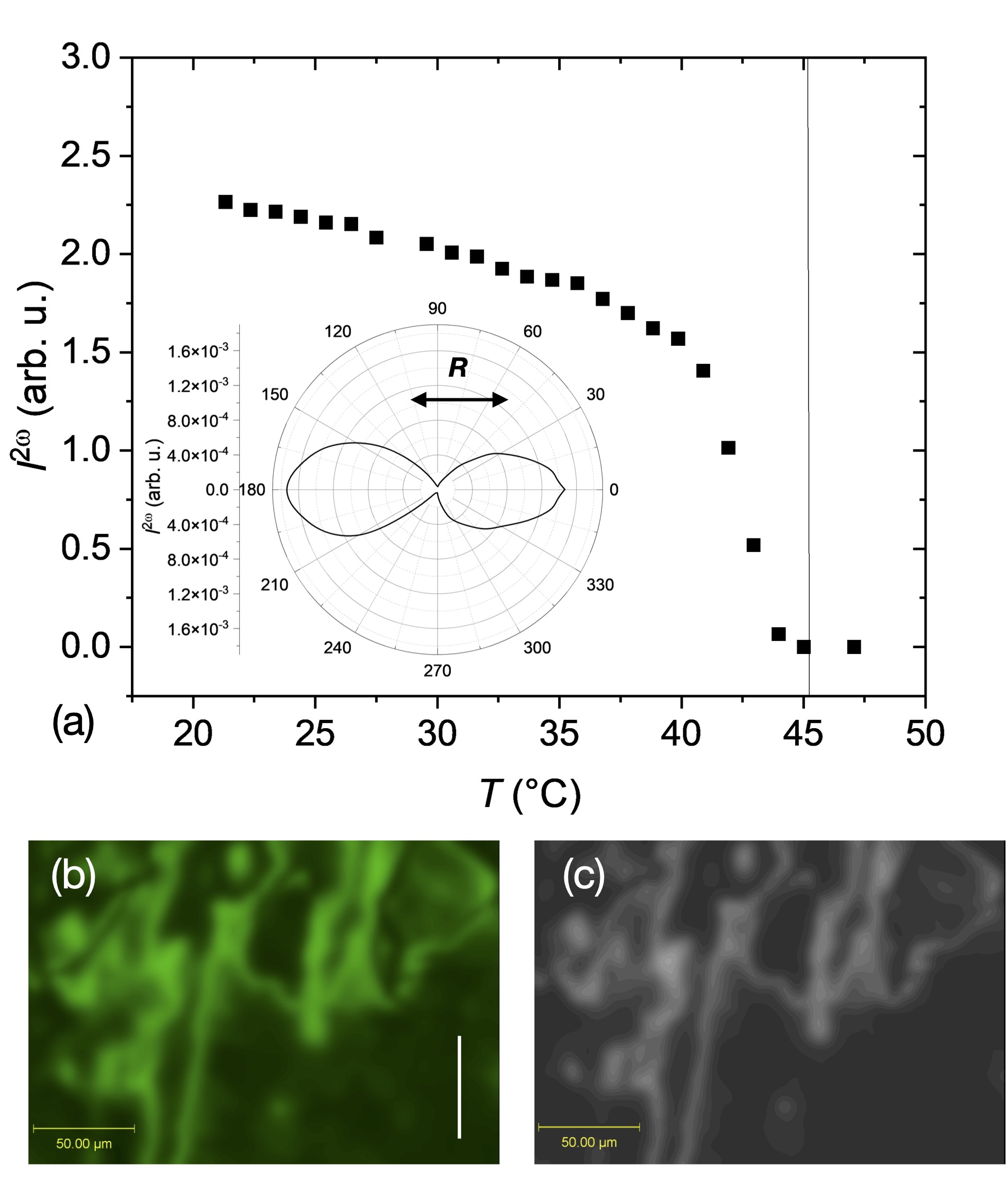}
  \caption{Second Harmonic Generation: (a) Temperature dependence of the $I^{2\omega}(T)$ of a sample confined in a cell with planar director anchoring. The inset shows the dependence of the SHG signal on the angle between the alignment direction $\mathbf{R}$ and the polarisation plane of the probing laser ($T=\qty{21}{\celsius}$). (b) Fluorescence microscopy image of a sample with untreated substrates and (c) the corresponding SHG microscopy image. The images are taken at $T=\qty{21}{\celsius}$ and the bar indicates the polarisation state of the probing laser.}
  \label{fig:shg}
\end{figure}

To investigate the polar nature of the N$_{\mathrm{F}}$ phase, we performed second harmonic generation (SHG) measurements. A strong, spontaneous SHG signal emerges as the system transitions into the N$_{\mathrm{F}}$ phase, with its intensity exhibiting a pronounced temperature dependence (Fig.~\ref{fig:shg}a). In planar cells with parallel rubbing, which ensures uniform director alignment, SHG efficiency reaches a maximum when the polarisation of the incident beam is parallel to the nematic director (see inset in Fig.~\ref{fig:shg}a). A similar response was observed in thin films of the N$_{\mathrm{F}}$ phase prepared on untreated substrates.

For direct visualisation of the director field, we employed the dichroic dye BTBP, which preferentially absorbs light polarised along the nematic director. Fig.~\ref{fig:shg}b shows the fluorescence pattern in a thin section of the sample, while Fig.~\ref{fig:shg}c presents the corresponding SHG microscopy image of the same region. The close similarity between the fluorescence and SHG textures confirms the parallel alignment of the nematic polar directors.

\subsection{\label{sec:dielectrics} Dielectric properties}

Dielectric spectroscopy (DS) was employed to investigate the polarisation dynamics across the isotropic and liquid crystalline phases of the studied compound. DS is a powerful technique for probing the relaxation of dipolar correlations and the emergence of collective dipolar motions, making it an indispensable tool for the characterisation of ferroelectric nematogens.

\begin{figure}[ht!]
\centering
 \includegraphics[width=0.8\columnwidth]{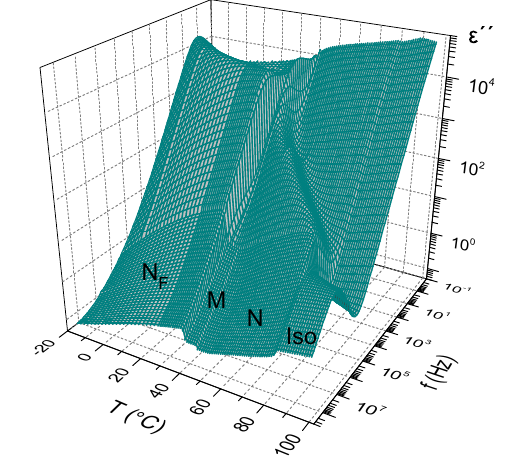}
  \caption{3D plot of the imaginary part of the complex dielectric permittivity ($\varepsilon''$) as a function of temperature and frequency measured in 50 \textmu m thick parallel plate capacitor with untreated gold-plated brass electrodes.}
  \label{fig:3D}
\end{figure}

The dielectric response was recorded over a wide frequency range ($\qty{0.1}{\hertz} -\qty{1}{\giga\hertz}$) and is presented in Fig.~\ref{fig:3D} as a three-dimensional plot of the imaginary part of the dielectric permittivity ($\varepsilon''$) as a function of temperature and frequency. In the isotropic phase, a single relaxation process is observed, with the frequency of maximum dielectric loss, $f_{\mathrm{max}}$, located around \qty{1e8}{\hertz}. Upon cooling into the nematic phase, a relaxation process emerges at lower $f_{\mathrm{max}}$, accompanied by a notable increase in dielectric strength. This finding indicates that the untreated gold-plated electrodes promote spontaneous homeotropic alignment of the nematic director, thereby enabling the measurement of the dielectric permittivity component parallel to the director. This behaviour is consistent with previous reports on prototypical ferroelectric nematogens, such as RM734 and DIO, where a strong tendency for out-of-plane director alignment was also observed in similar sample cells with untreated gold-plated electrodes~\cite{Sebastian.2020, Erkoreka.2023, erkoreka2023collective}.
Although the director alignment is determined implicitly by the use of untreated electrodes, this configuration ensures that insulating alignment layers do not contribute to the total measured capacitance. This is a crucial consideration for interpreting the dielectric permittivity in the N$_{\mathrm{F}}$ phase~\cite{Brown.2021ysv, Clark.2024, Adaka.2024, Vaupotic.2023, matko2024interpretation}.

\begin{figure}[ht!]
\centering
 \includegraphics[width=0.8\columnwidth]{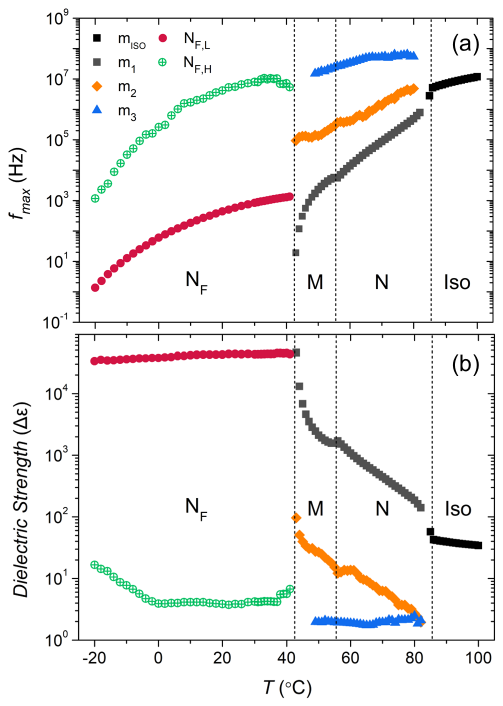}
  \caption{(a) Temperature dependence of the relaxation frequency maxima $\left(f_{max}\right)$ and (b) the dielectric strengths ($\Delta\varepsilon$) of the recorded relaxation mechanisms, obtained through fitting of the dielectric spectra with Eq.~\ref{eq:HN}.}
  \label{fig:Arrhenius}
\end{figure}

The deconvolution of the dielectric spectra was performed by simultaneously fitting the real, $\varepsilon'(\omega)$, and imaginary, $\varepsilon''(\omega)$, components of the dielectric permittivity using Eq.~\ref{eq:HN}. Representative examples of the deconvoluted spectra for the studied material are shown in Fig.~\ref{sfig:DS-terms}. The temperature dependence of the relaxation frequency maxima ($f_{\mathrm{max}}$) and the corresponding dielectric strengths ($\Delta\varepsilon$) for each fitted relaxation mode are presented in Fig.~\ref{fig:Arrhenius}a and Fig.~\ref{fig:Arrhenius}b, respectively.

In the isotropic phase, the spectra were well described by a single relaxation mode ($m_{\mathrm{iso}}$) of Havriliak–Negami type, with a maximum loss frequency on the order of $10^7$~Hz and a dielectric strength of approximately ${\Delta\varepsilon}_{m,\mathrm{iso}} \approx 40$. This relaxation process is attributed to the overall reorientational motions of dipolar units in the isotropic environment and follows an Arrhenius temperature dependence:
\[
f_{\mathrm{max}}(T) = f_0 \exp\left(-\frac{E_{\mathrm{A}}}{k_{\mathrm{B}} T}\right),
\]
\noindent with an activation energy $E_{\mathrm{A}} \approx \qty{65}{\kilo\joule\per\mol}$, consistent with values reported for conventional nematogens~\cite{goodby2014handbook}.

In the N and M phases, three distinct relaxation mechanisms are identified, denoted $m_1$, $m_2$, and $m_3$ in order of increasing relaxation frequency (Fig.~\ref{fig:Arrhenius}a). The most prominent in terms of dielectric strength is the $m_1$ mode, which exhibits a nearly Debye-like character, with $\alpha_1 \approx 0.97$. The intermediate-frequency mode $m_2$ also displays a close-to-Debye shape, featuring symmetric broadening of the relaxation time distribution ($\alpha_2 \approx 0.9$–$0.7$). In contrast, the high-frequency $m_3$ mode shows an asymmetric distribution, characterised by $\alpha_3 \approx 0.8$ and $\beta_3 \approx 0.6$.

The frequency of maximum loss of the $m_1$ mode, $f_{\mathrm{max},m1}$, progressively decreases upon cooling in the N phase, while its dielectric strength, $\Delta\varepsilon_{m1}$, increases sharply. This mode is attributed to reorientations of the longitudinal component of the dipolar groups about the short molecular axis, analogous to the $\omega_1$ relaxation predicted by the Nordio–Rigatti–Segre theory for uniaxial nematics composed of rigid dipolar molecules~\cite{luigi1973dielectric}. The observed decrease in $f_{\mathrm{max},m1}$ with decreasing temperature reflects the increasing orientational order and viscosity of the system.

However, two key features distinguish the $m_1$ process from the classical $\omega_1$ relaxation of conventional nematogens. First, the dielectric strength increases dramatically upon cooling, reaching values of $\Delta\varepsilon_{m1} \approx 2000$ near the M–N$_{\mathrm{F}}$ phase transition. Second, while $f_{\mathrm{max},m1}$ follows Arrhenius temperature dependence - as expected for this type of molecular reorientation - the associated activation energy is significantly higher, with $E_{\mathrm{A},m1} \approx \qty{170}{\kilo\joule\per\mol}$. This value is nearly twice than reported for similar reorientations in conventional nematogens~\cite{goodby2014handbook}, and substantially higher than that observed for the corresponding process in DIO~\cite{erkoreka2023collective}.

Within the antiferroelectric M phase, and upon approaching the N$_{\mathrm{F}}$ transition, the $m_1$ mode exhibits soft-mode-like behaviour—characterised by a strong reduction in $f_{\mathrm{max},m1}$ accompanied by a divergent increase in dielectric strength. This trend is consistent with observations in DIO for the analogous relaxation mode~\cite{erkoreka2023collective}.

These findings suggest that the effective mean squared dipole moment along the nematic director is strongly influenced by short-range orientational dipolar correlations, quantified by the Kirkwood correlation factor $g_1$ \cite{kirkwood1939dielectric, bordewijk1974extension}. Although these correlations are short-ranged, they must be considered to rationalise both the magnitude and temperature dependence of the dielectric permittivity—particularly in anisotropic polar fluids—as recently demonstrated in the case of dimeric liquid crystals~\cite{zavvou2023dipole}.

Interestingly, weaker collective dipolar reorientations have also been reported in the nematic phase of the non-N$_{\mathrm{F}}$-forming material RM734-CN~\cite{Mandle.2021}. In the present system, the pronounced reduction of $f{_{\mathrm{max},m1}}$, accompanied by a sharp increase in ${\Delta\varepsilon}_{m1}$, can be attributed to collective fluctuations of dipolar clusters of molecules with parallel mutual alignment along the nematic director.

The deviation of the $m_1$ mode from Arrhenius behaviour in the M phase - marked by a critical slowing down of $f_{\mathrm{max},m1}$ - further suggests that the strength and extent of these dipolar correlations increase significantly in the intermediate antiferroelectric phase. These correlations appear to grow continuously until they become long-ranged at the transition to the N$_{\mathrm{F}}$ phase.

A few degrees below the I–N transition, the contribution of the intermediate $m_2$ relaxation process was deconvoluted through simultaneous fitting of the real and imaginary parts of the dielectric permittivity, along with assessment of the fit quality in the derivative $d(\varepsilon’)/d(\log f)$ (see Fig.~\ref{sfig:DS-terms}). Notably, this represents the first time that the relaxation dynamics of the $m_2$ process have been deconvoluted across nearly the entire temperature range of the nematic phase. In contrast, the analogous low- and intermediate-frequency processes in DIO~\cite{Brown.2021ysv, erkoreka2023collective, yadav2023two} and RM734~\cite{Sebastian.2020, Mandle.2021} typically exhibit strong spectral overlap deep into the N phase. The clear time-scale separation between $m_2$ and $m_1$ in the present system is likely facilitated by the higher activation energy of the latter.

As shown in Fig.~\ref{fig:Arrhenius}a, the relaxation frequency maximum of the $m_2$ mode follows an Arrhenius temperature dependence in the N phase, with an activation energy of $E_{\mathrm{A},m2} \approx \qty{110}{\kilo\joule\per\mol}$. The corresponding dielectric strength, $\Delta\varepsilon_{m2}$, increases upon cooling but remains nearly two orders of magnitude lower than $\Delta\varepsilon_{m1}$ (Fig.~\ref{fig:Arrhenius}b). As the system approaches the M–N$_{\mathrm{F}}$ phase transition, the growth of polar order appears to influence the $m_2$ process, as evidenced by the increase in $\Delta\varepsilon_{m2}$, while the softening of its relaxation frequency was also observed in DIO~\cite{erkoreka2023collective}.

These findings suggest that the $m_2$ process may be associated with individual dipolar reorientations, at least at higher temperatures within the N phase. Upon cooling, a clear time-scale separation emerges between the slower collective dipolar fluctuations of the $m_1$ mode and the faster, more localized motions of $m_2$, possibly occurring within dipolar clusters. Based on the observed activation energy and growing dielectric strength, the $m_2$ mode may involve reorientation of dipoles about the short molecular axis. However, this interpretation is somewhat counterintuitive, as the relaxation frequency does not slow significantly near the transition, suggesting that some form of precessional or hindered rotational motion may also be involved.

Yadav et al.~\cite{yadav2023two} proposed that the intermediate mode in the high-temperature nematic phase of DIO is collective in nature, originating from the segregation of chiral conformers into domains with the same sense of chirality, with correlation lengths increasing upon cooling. The precise origin of the $m_2$ mode in the present system remains unclear and warrants further investigation, which is beyond the scope of this work.

The $m_3$ mode, which is the weakest in terms of dielectric strength, is recorded in the high-frequency region of the spectra. This process remains nearly temperature independent up to the middle of the N phase ($E_{A,m3}\approx12 \, \text{kJ/mol}$), while a slow reduction of the relaxation frequency is observed on further cooling. Based on the magnitude and the temperature dependence of its dielectric strength this process can be assigned to individual fluctuations of the projections of the dipolar groups around the long molecular axis. This is also in line with the weak temperature variation of the relaxation time, as such motions are not hindered by the nematic potential. These trends are in agreement with those reported both for RM734 ~\cite{Sebastian.2020, Mandle.2021}  and DIO ~\cite{erkoreka2023collective} .

Concerning the N$_{\mathrm{F}}$ phase, two relaxation mechanisms are recorded denoted as $N_{\mathrm{F,L}}$ and $N_{\mathrm{F,H}}$ following the notation proposed by Erkoreka et al. ~\cite{erkoreka2023collective, erkoreka2024molecular}, for the low- and high-frequency processes recorded in UUQU-4-N and DIO. The $N_{\mathrm{F,L}}$ mode, a nearly Debye-type process located around \qty{1}{\kilo\hertz}, is distinguished by the huge and relatively temperature-independent dielectric strength in the order of $10^4$. This process can be identified as a Goldstone (phason) mode according to Vaupoti\v{c} et al. ~\cite{Vaupotic.2023}, due to coupled fluctuations of director and polarisation or as a Polarisation-External Capacitance-Goldstone (PCG) mode proposed by Clark et.al. ~\cite{Clark.2024}. The relaxation frequency of $N_{\mathrm{F,L}}$ exhibits a negligible temperature dependence close the transition, while on further cooling below room temperature the reduction of the relaxation frequency is due the increase in material viscosity. The high frequency mode $N_{\mathrm{F,H}}$ exhibits a Havriliak-Negami shape with a relatively small strength and a slowly decreasing relaxation frequency. This process could be assigned to molecular reorientations around the long molecular axis, i.e. the evolution of the $m_3$ mode in the N$_{\mathrm{F}}$ phase. Another possibility is that $N_{\mathrm{F,H}}$ corresponds to the high frequency phason mode predicted by the Continuous Phenomenological Model of Vaupoti\v{c} et al. ~\cite{Vaupotic.2023}.

However, in light of recent studies concerning the interpretation of the dielectric properties of ferroelectric nematogens, a rather unanimous conclusion is that the giant capacitance values measured in the N$_{\mathrm{F}}$ phase are apparent. The way that these apparent values are connected to the intrinsic material properties, i.e. the capacitance and resistance of the N$_{\mathrm{F}}$ layer, depends on the exact choice of the equivalent circuit for analysing the dielectric spectra. Clark et al. ~\cite{Clark.2024} have suggested that the block reorientation of spontaneous polarisation renders the N$_{\mathrm{F}}$ phase effectively conductive. This way the measured capacitance corresponds to the capacitance of the interfacial layers between the electrodes and the bulk material, either the polyimide insulating layers or the thin layers of the material itself. On the other hand, Matko et al. ~\cite{matko2024interpretation} and Vaupoti\v{c} et al. ~\cite{vaupotivc2025ferroelectric} proposed the “high-$\varepsilon$” model, arguing that the analysis of the complete equivalent circuit in both bare and treated electrodes results in a resistivity of the N$_{\mathrm{F}}$ layer being of a similar magnitude to conventional nematogens and that the capacitance of the N$_{\mathrm{F}}$ phase is even higher than the measured apparent values. While the analysis of dielectric spectra through an appropriate equivalent circuit is of significant interest, it falls outside the scope of this study.

\subsection{\label{sec:mechanics} Mechanical properties}

Reorientation of the uniformly aligned nematic director in a sandwich cell under a magnetic field is governed by the balance between magnetic and elastic torques. When the applied magnetic field exceeds a threshold value, the initially undisturbed director orientation becomes unstable — a phenomenon known as the \textit{magnetic Fr\'eedericksz transition} (mFT)~\cite{deGennes:1995vg, Freedericksz.1927} . An analogous effect occurs under an electric field, termed the \textit{electric Fr\'eedericksz transition} (eFT), where a dielectric torque acts upon the director.
\begin{figure}[h]
\centering
  \includegraphics[width=0.8\columnwidth ]{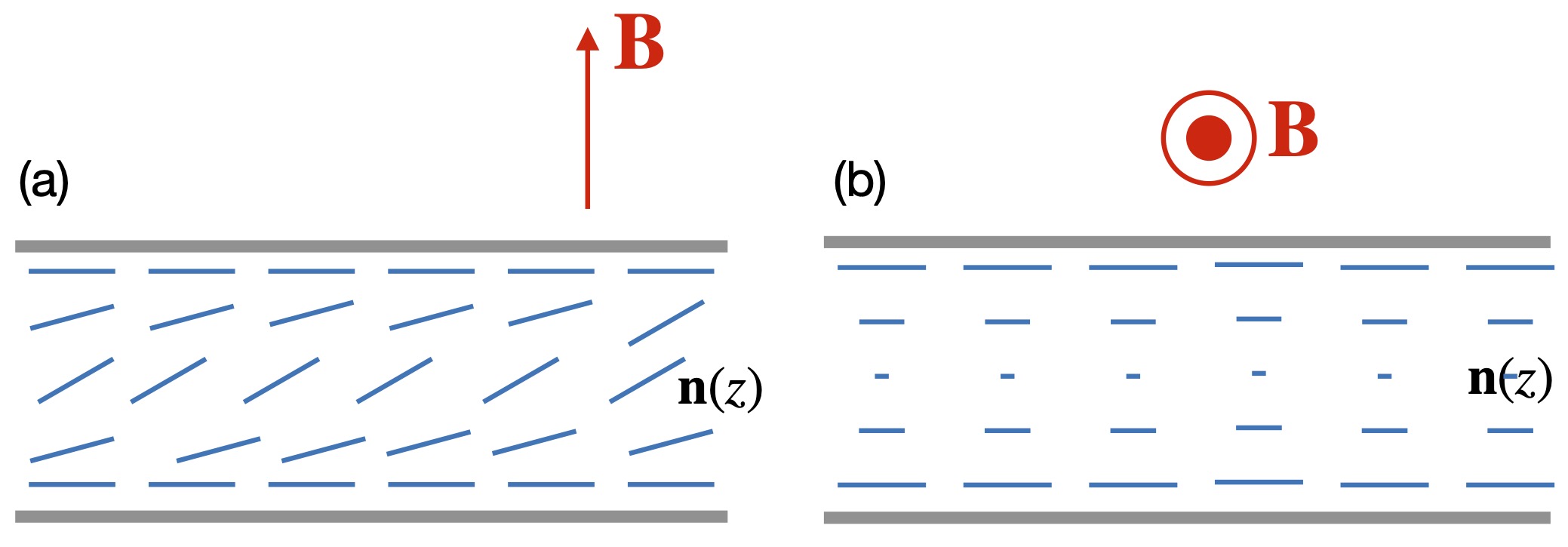}
  \caption{Schematic representation of the measurement geometries for the splay (a) and twist (b) elastic constants using the magnetic Fr\'eedericksz transition. $\mathbf{n}$ and ($\mathbf{B}$) are the nematic director and the magnetic field, respectively. Strong planar anchoring condition is assumed.  }
    \label{fig:geometry}
\end{figure}

In both cases, the critical fields \( B_c=\pi/d\sqrt{\mu_0K_i/\chi_{\mathrm{m,a}}} \) and \( E_c=\pi/d\sqrt{K_i/\varepsilon_0\chi_{\mathrm{e,a}}}  \) are determined by the Frank elastic constants $K_i$ and the corresponding anisotropies: diamagnetic (\( \chi_{\mathrm{m,a}} \)) for mFT, and dielectric (\( \chi_{\mathrm{e,a}} \)) for eFT. Therefore, measurements of the Fr\'eedericksz threshold fields offer a reliable method for extracting the Frank elastic constants in nematic systems.

Magnetic Fr\'eedericksz transitions are generally more straightforward to interpret, as magnetic fields do not couple directly to electric charges in the material, eliminating the influence of conductivity and the flexoelectric polarisation~\cite{Zavvou.2022gmk}. The splay transition is induced when a magnetic field is applied perpendicular to the substrate in a homogeneously aligned nematic cell, while a twist transition is observed when the field is applied parallel to the substrate and perpendicular to the undisturbed director (Fig.~\ref{fig:geometry}). A limitation of magnetic field measurements, however, is the need for accurate knowledge of the diamagnetic anisotropy, which is often challenging to determine directly.

\begin{figure}[h]
\centering
  \includegraphics[width=0.7\columnwidth ]{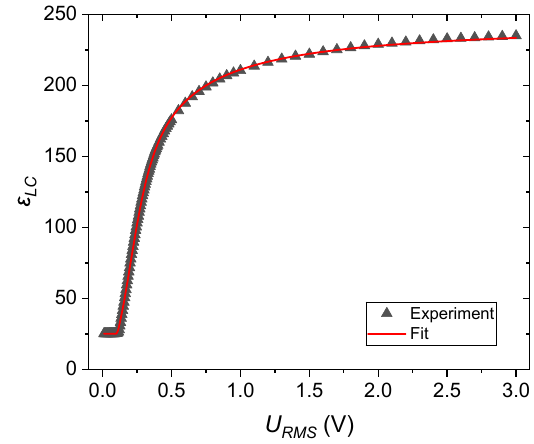}
  \caption{Voltage dependence of the dielectric permittivity of the LC layer measured in the N phase at T=78.5$^\circ$C and f=1.5 kHz. The solid line represents the fit using ~\cref{eq:DEU_1,eq:DEU_2}. }
    \label{fig:CV}
\end{figure}

\begin{figure}[h]
\centering
  \includegraphics[width=0.6\columnwidth ]{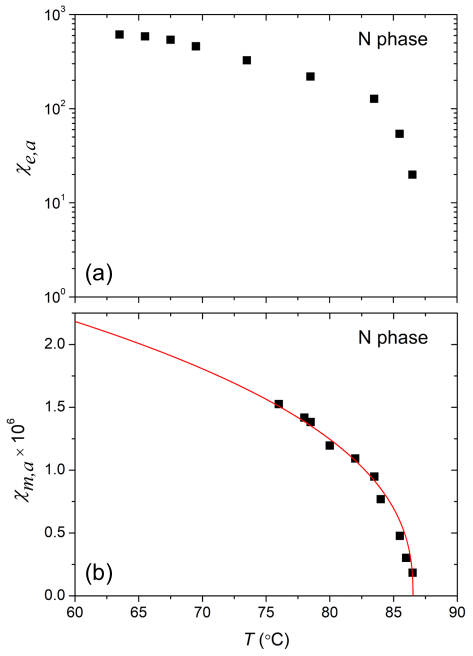}
  \caption{Temperature dependences of (a) dielectric anisotropy (\( \chi_{\mathrm{e,a}} \))  and (b) diamagnetic anisotropy (\( \chi_{\mathrm{m,a}} \)) in the N phase. The red line represents the Haller fit.}
    \label{fig:deltas}
\end{figure}

\begin{figure}[h]
\centering
  \includegraphics[width=0.5\columnwidth ]{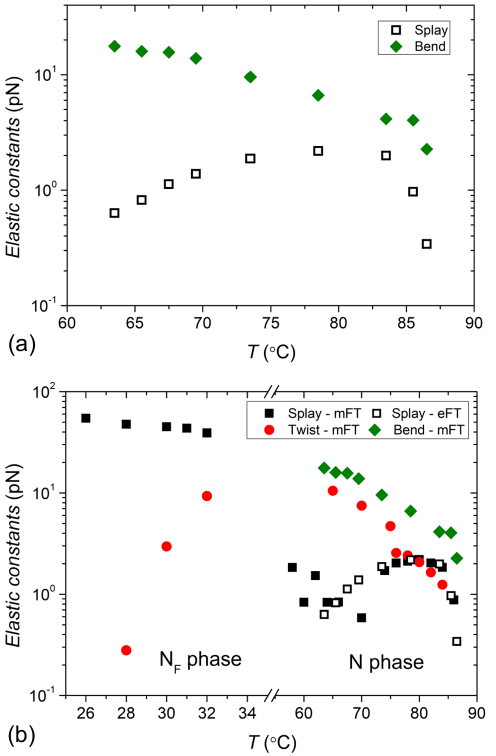}
  \caption{Elastic constants in the N and N$_{\mathrm{F}}$ phases: (a) the splay (\( K_{11} \)) and bend (\( K_{33} \)) elastic constants in the N phase  determined by eFT and fitting the dielectric permittivity $\varepsilon_{LC}(U)$ with \cref{eq:DEU_1,eq:DEU_2}. (b) Comparison of the elastic constants determined using the mFT and eFT.}
    \label{fig:ElConst_ALL}
\end{figure}

In this study, we first determined the dielectric permittivity in the N phase as a function of the applied AC voltage across the cell. The recalculated $\varepsilon_{LC}(U)$ curves were fitted using Deuling's solution~\cite{Deuling:2007kj} for the director equilibrium equations, with a representative example depicted in Fig.~\ref{fig:CV}. The threshold voltage obtained in the N phase is in the order of $U_c=0.1 V$, much smaller than in conventional nematics, and decreases upon cooling (see Fig.~\ref{sfig:Thresholds}a). The dielectric permittivity $\varepsilon_{LC}$ saturates at high fields, allowing the determination of the dielectric anisotropy through Deuling fitting. Although, the effect of the PI layers is discussed in the literature with respect to their effect on the measured permittivity in the N$_{\mathrm{F}}$ phase, we find that due to the high dielectric anisotropy of the N phase, the $\varepsilon_{LC}(U)$ can be significantly underestimated unless the contribution of the PI layers to the total measured sample capacitance is taken into account (see Fig.~\ref{sfig:CV_examples}). This is especially important when measurements are conducted in thin cells with an increased $d_{\mathrm{PI}} / d$ ratio, that would result in an effective reduction of the measured permittivity and the splay elastic constant with decreasing the sample thickness. Regarding the fitting process, Deuling’s solution is valid sufficiently far from the nematic–modulated (N–M) phase transition (Fig.~\ref{fig:ElConst_ALL}a).  Closer to the transition, significant deviations of \( \varepsilon_{LC}(U) \) from the model suggest an increasing role of flexoelectric coupling between the director and the electric field. Near the transition to the M phase, only the splay elastic constant could be reliably estimated from the threshold voltage.

The dielectric anisotropy (\( \chi_{\mathrm{e,a}} \)) extracted through eFT studies in the high-temperature nematic phase is presented in Fig.~\ref{fig:deltas}a. \( \chi_{\mathrm{e,a}} \) increases markedly on cooling, exceeding the values obtained in conventional nematics by at least one order of magnitude. The determination of \( \chi_{\mathrm{e,a}} \) and threshold voltage ($U_c$) through eFT (Fig.~\ref{sfig:Thresholds}a), along with the critical magnetic field strength ($B_c$) through mFT (Fig.~\ref{sfig:Thresholds}b) at the same temperatures, allows for the estimation of the temperature dependence of the diamagnetic anisotropy $\chi_{\mathrm{m,a}} = \varepsilon_0 \mu_0 \, \chi_{\mathrm{e,a}} {\left( U_c/d B_c \right)}^2$, shown in Fig.~\ref{fig:deltas}b. A Haller-type fit was applied in the high-temperature nematic phase to extrapolate values of the diamagnetic anisotropy into the ferroelectric nematic (N$_{\mathrm{F}}$) region (Fig.~\ref{fig:deltas}b)~\cite{Haller1975}:

\[
\chi_{\mathrm{m,a}} = \chi_{\mathrm{m,a0}} \left(1 - \frac{T}{T^*}\right)^\beta
\]
\noindent where $\chi_{\mathrm{m,a0}}$ and $\beta$ are empirical constants and $T^*$ is an extrapolated temperature. The obtained fitting parameters are $\chi_{\mathrm{m,a0}}=(6\pm1)\times 10^{-6}$, $T^*=(359.8\pm0.3)K$ and $\beta=(0.40\pm0.06)$. The Haller-type behaviour of the orientational order parameter across the whole range of N and N$_{\mathrm{F}}$ phases was established by birefringence measurements.

The temperature dependencies of the splay (\( K_{11} \)), bend (\( K_{33} \)), and twist (\( K_{22} \)) elastic constants are shown in Fig.~\ref{fig:ElConst_ALL}a,b. In the nematic phase, \( K_{11} \) exhibits a pronounced decrease with decreasing temperature, especially as the system approaches the transition to the modulated phase M. In contrast, both \( K_{22} \) and \( K_{33} \) increase significantly. Two key effects may explain this behaviour: First, the flexoelectric coupling becomes increasingly important. The Landau–de Gennes (LdG) free energy for a ferroelectric nematic includes expansions in both the orientational order parameter \( Q \) and the polar order parameter \( P \), containing various coupling terms~\cite{Paik.2024,Shamid:2013ib,Mertelj.2018}. The paraelectric–ferroelectric transition is primarily driven by terms involving powers of \( P \). Notably, the term quadratic in \( P \), with a temperature-dependent coefficient \( t \), drives the system toward a ferroelectric state characterised by spontaneous splay deformation~\cite{Shamid:2013ib,Mertelj.2018}. Even in the nematic phase (\( t > 0 \)), this term, in combination with the flexoelectric coupling, leads to a reduction of the effective splay constant \( K_{\text{eff}} = K_{11} - \frac{\gamma^2}{t} \), proportional to the square of the flexoelectric coefficient \( \gamma \).

Second, the emergence of modulated, smectic-like order in the M phase suppresses bend elasticity, contributing to the increasing bend rigidity. In contrast, twist deformation remains largely unaffected, as the modulation wave-vector is orthogonal to the director.

In the N$_\mathrm{F}$ phase, the behaviour changes significantly. The splay elastic constant exhibits a sharp increase by nearly an order of magnitude. As demonstrated in our previous work on an example of a compound with a single N$_\mathrm{F}$ phase,  this enhancement is attributed to electrostatic contributions to the system's free energy ~\cite{Zavvou.2022gmk}. The role of the electrostatic contributions have been discussed in various publications ~\cite{Paik.2024,Clark.2024,Caimi.2023t5j,Mathe.2025} . Splay deformations of the director generate bound charges $\rho_{\mathrm{b}}=-\nabla\cdot\mathbf{P}$, introducing an energetically unfavourable electrostatic component. The bound charges interact via screened electrostatic interactions, contributing the energy 
\[
U_{\mathrm{ES}}=\frac{1}{8 \pi \epsilon \epsilon_0} \iint \frac{\rho(\mathbf{r}) \rho\left(\mathbf{r}^{\prime}\right) e^{-\kappa\left|\mathbf{r}-\mathbf{r}^{\prime}\right|}}{\left|\mathbf{r}-\mathbf{r}^{\prime}\right|} \mathrm{d} \mathbf{r}^{\prime} \mathrm{d} \mathbf{r}
\]
\noindent where $\kappa$ is the inverse Debye screening length~\cite{Paik.2024} (see also~\ref{SI:math}). Assuming that $\mathbf{P}(\mathbf{r})=P_0\mathbf{n}(\mathbf{r})$, it is easy to show that splay elastic term has the same form as the electrostatic term yielding the effective splay elastic constant
\[
K_{11}^{\text{eff}}(k) = K_{11} + \frac{P_0^2}{\varepsilon \varepsilon_0 (k^2 + \kappa^2)},
\]
\noindent which is for the long wavelengths limit gives $K_{11}^{\text{eff}}(k) \approx K_{11} + P_0^2 \lambda_D^2 / (\varepsilon \varepsilon_0)$ with the Debye length $\lambda_D=1/\kappa$. 
Substituting the typical values of $P_0=\qty{6}{\micro\coulomb\per\centi\meter\squared}$, bare liquid crystal dielectric permittivity $\varepsilon=100$, and a typical Debye length of $\lambda_D\approx\qty{100}{\nano\meter}$ giving the correction \qty{400}{\pico\newton}. Although this figure has a similar order of magnitude as observed in the experiment, the discrepancy can be attributed to the absence of the exact values for the bare dielectric constant and the Debye screening length.

The twist elastic constant exhibits significant softening in the N$_\mathrm{F}$ phase (Fig.\ref{fig:ElConst_ALL}b). This twist instability may be understood as a result of competing elastic and electrostatic forces. In systems with degenerate anchoring, Coulomb interactions promote the formation of ambidextrous twist configurations, as predicted by Khachaturyan in~\cite{khachaturyan1975development}. Similar phenomena have been observed in other ferroelectric systems, such as bent-core nematics, where splay rigidity influences the formation of topological defects and inversion walls in freely suspended films.

Ferroelectric nematics, with their true three-dimensional fluidity, offer a unique opportunity to study the interplay of elastic and electrostatic forces. This interplay gives rise to rich director field configurations, complex defect structures, and intriguing self-assembly behaviours under applied electric fields.

In conclusion, we have measured and compared the Frank elastic constants in a material exhibiting both paraelectric and ferroelectric nematic phases, primarily via magnetic Fr\'eedericksz transition. Our findings confirm the softening of splay elasticity in the paraelectric phase and its subsequent hardening in the N$_{\mathrm{F}}$ phase, which we attribute to the electrostatic contributions arising from spontaneous polarisation.

\section*{Acknowledgments}
We acknowledge the financial support of Deutsche Forschungsgemeinschaft (Projects NA1668/1-3 and ER 467/8-3). The authors thank Dr. Nerea Sebasti\'an and Dr. Alenka Mertelj (Jo\^zef Stefan Institute, Slovenia), Dr. Josu Martinez-Perdiguero (University of the Basque Country UPV/EHU, Spain) and Prof. Mikhail Osipov (University of Strathclyde, UK) for fruitful and motivating discussion. E. Z. acknowledges financial support from the Hellenic Foundation for Research and Innovation(HFRI PhD Fellowship grant, Fellowship Number: 809096).

\bibliographystyle{elsarticle-num} 

\newpage

\bigbreak

\clearpage

\section*{Supplementary information}
\renewcommand{\thefigure}{S\arabic{figure}}
\setcounter{figure}{0}
\renewcommand{\thesubsection}{S\arabic{subsection}}
\setcounter{subsection}{0}

\subsection{Dielectric properties}\label{SI-Diel}
\bigbreak
\begin{figure*}[h!]
\centering
  \includegraphics[width=0.75\textwidth]{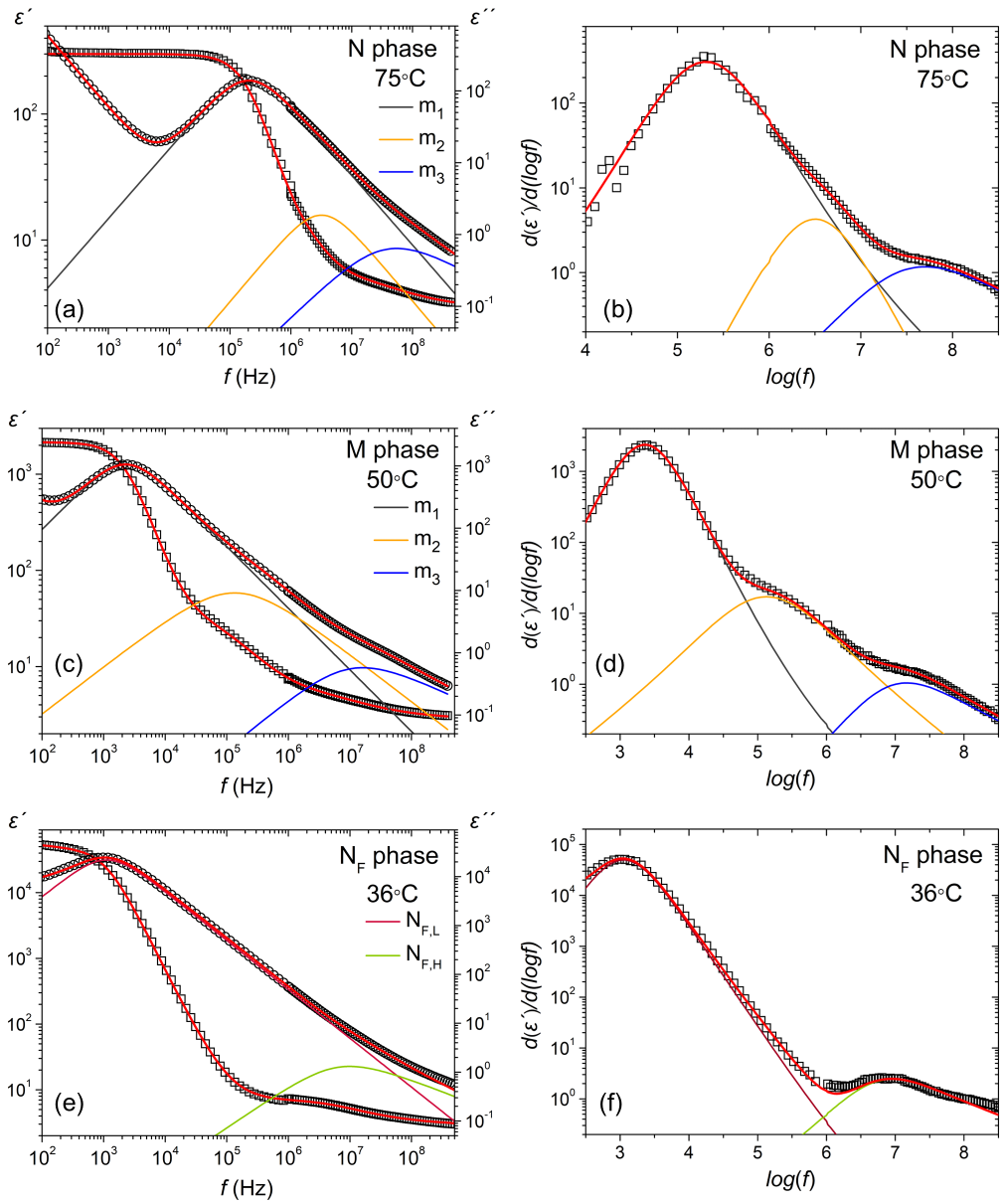}
  \caption{(Left) Representative examples of the deconvoluted isothermal curves of the real $\varepsilon'$ (squares) and imaginary $\varepsilon''$ (circles) parts of the dielectric permittivity of the studied material versus frequency in (a) the N phase at $75\,^\circ\mathrm{C}$, (b) the M phase at $50\,^\circ\mathrm{C}$ and (e) the N$_{\mathrm{F}}$ phase at $30\,^\circ\mathrm{C}$. Open symbols correspond to experimental data, while solid lines are fits according to Havriliak-Negami equation. (Right) The derivative of the real part of the dielectric permittivity $d(\varepsilon')⁄d(logf)$ versus $logf$. }
    \label{sfig:DS-terms}
\end{figure*}

\subsection{Mechanical properties}\label{SI-FT}
\bigbreak
\begin{figure*}[ht]
\centering
  \includegraphics[width=0.75\textwidth]{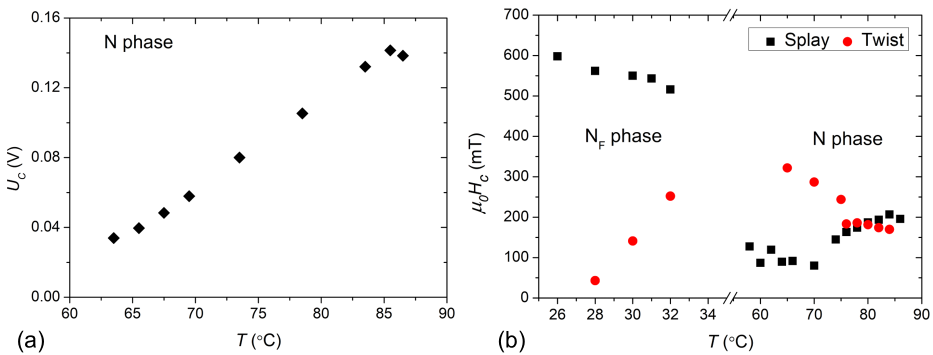}
  \caption{Temperature dependence of (a) the threshold voltage ($U_{c}$) of the onset of splay Fr\'eedericksz transition obtained through capacitance measurements in the N phase and (b) the critical magnetic field strength ($\mu_0H_{c}$) of the onset of splay (squares) and twist (circles) Fr\'eedericksz transition obtained through magneto-optical measurements in the N and N$_{\mathrm{F}}$ phases of the studied compound.}
    \label{sfig:Thresholds}
\end{figure*}

\begin{figure*}[ht]
\centering
  \includegraphics[width=\textwidth]{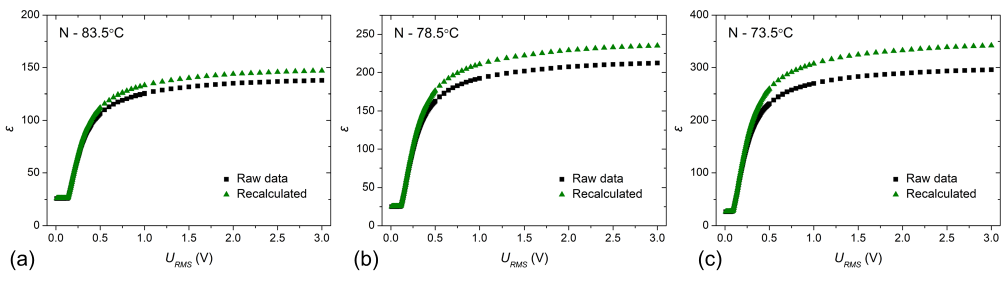}
  \caption{Voltage dependence of the dielectric permittivity at various temperatures across the N phase. Square symbols represent the raw measured dielectric permittivity, defined as $\varepsilon_{EQ}=C_{EQ}/C_{empty}$, while triangles correspond the recalculated permittivity of the LC layer, defined as $\varepsilon_{LC} = \varepsilon_{EQ} \varepsilon_{PI}/\left[\varepsilon_{PI} - 2\varepsilon_{EQ}(d_{PI}/d)\right]$. }
    \label{sfig:CV_examples}
\end{figure*}

\subsection{Electrostatic splay-stiffening}\label{SI:math}

We consider the contribution to the free energy arising from screened Coulomb interactions between bound (polarisation) charges. This contribution is expressed as:
\[
F_\rho = \frac{1}{8 \pi \varepsilon \varepsilon_0} \iint \rho(\mathbf{r}) \rho(\mathbf{r}{\prime}) \frac{e^{-\kappa |\mathbf{r} - \mathbf{r}{\prime}|}}{|\mathbf{r} - \mathbf{r}{\prime}|} \, d\mathbf{r}{\prime} \, d\mathbf{r},
\]
where $\rho = -\nabla \cdot \mathbf{P}$ denotes the bound charge density associated with the polarisation field $\mathbf{P}$, and $\kappa$ is the inverse Debye screening length. This formulation assumes that polarisation splay generates an electrostatic potential given by:
\[
\phi(\mathbf{r}) = \frac{1}{4 \pi \varepsilon \varepsilon_0} \int \frac{e^{-\kappa |\mathbf{r} - \mathbf{r}{\prime}|}}{|\mathbf{r} - \mathbf{r}{\prime}|} \rho(\mathbf{r}{\prime}) \, d\mathbf{r}{\prime},
\]
where $\phi(\mathbf{r})$ is the potential at the point $\mathbf{r}$.

To simplify the analysis, we switch to Fourier space, using:
\[
\rho(\mathbf{r}) = \int \tilde{\rho}(\mathbf{k}) e^{i \mathbf{k} \cdot \mathbf{r}} \frac{d\mathbf{k}}{(2\pi)^3}, \quad
\phi(\mathbf{r}) = \int \tilde{\phi}(\mathbf{k}) e^{i \mathbf{k} \cdot \mathbf{r}} \frac{d\mathbf{k}}{(2\pi)^3},
\]
where $\tilde{\rho}(\mathbf{k})$ and $\tilde{\phi}(\mathbf{k})$ are the Fourier transforms of the charge density and potential, respectively.

Substituting into the free energy expression and simplifying:
\[
F_\rho = \frac{1}{2} \int \left[ \int \tilde{\rho}(\mathbf{k}) e^{i \mathbf{k} \cdot \mathbf{r}} \frac{d\mathbf{k}}{(2\pi)^3} \right]
\left[ \int \tilde{\phi}(\mathbf{q}) e^{i \mathbf{q} \cdot \mathbf{r}} \frac{d\mathbf{q}}{(2\pi)^3} \right] d\mathbf{r},
\]
we arrive at the compact form:
\[
F_\rho = \frac{1}{2 \varepsilon \varepsilon_0} \int \frac{|\tilde{\rho}(\mathbf{k})|^2}{k^2 + \kappa^2} \frac{d\mathbf{k}}{(2\pi)^3}.
\]
Expressing the bound charge in terms of polarisation, $\tilde{\rho}(\mathbf{k}) = -i \mathbf{k} \cdot \tilde{\mathbf{P}}(\mathbf{k})$, the electrostatic free energy becomes:
\[
F_\rho = \frac{1}{2 \varepsilon \varepsilon_0} \int \frac{|\mathbf{k} \cdot \tilde{\mathbf{P}}(\mathbf{k})|^2}{k^2 + \kappa^2} \frac{d\mathbf{k}}{(2\pi)^3}.
\]
The elastic free energy contribution from splay deformations of the director field $\mathbf{n}$ is given by:
\[
F_\text{splay} = \frac{1}{2} K_1 \int (\nabla \cdot \mathbf{n})^2 \, d\mathbf{r},
\]
which in Fourier space becomes:
\[
F_\text{splay} = \frac{1}{2} K_1 \int |\mathbf{k} \cdot \tilde{\mathbf{n}}(\mathbf{k})|^2 \, \frac{d\mathbf{k}}{(2\pi)^3}.
\]
In the ferroelectric nematic (N$_\mathrm{F}$) phase, we assume that the polarisation is aligned with the director, i.e., $\mathbf{P} = P_0 \mathbf{n}$. This allows us to combine the electrostatic and splay elastic terms into a total free energy:
\[
F_\text{total} = \frac{1}{2} \int |\mathbf{k} \cdot \tilde{\mathbf{n}}(\mathbf{k})|^2 \left[ K_1 + \frac{P_0^2}{\varepsilon \varepsilon_0 (k^2 + \kappa^2)} \right] \frac{d\mathbf{k}}{(2\pi)^3},
\]
where the expression in brackets defines an effective splay elastic constant:
\[
K_\mathrm{eff}(k) = K_1 + \frac{P_0^2}{\varepsilon \varepsilon_0 (k^2 + \kappa^2)}.
\]
\end{document}